\documentclass[aps, prd, amsmath, floats, floatfix, twocolumn, nofootinbib, superscriptaddress]{revtex4-1}
\usepackage{graphicx}
\usepackage{color}
\usepackage{latexsym}

\begin{document}

\title{Accretion disk around the rotating Damour-Solodukhin wormhole}

\author{R.Kh. Karimov}
\affiliation{Zel'dovich International Center for Astrophysics, Bashkir State Pedagogical University, 3A, October Revolution Street, Ufa 450008, RB, Russia}

\author{R.N. Izmailov}
\email[Corresponding author: ]{izmailov.ramil@gmail.com}
\affiliation{Zel'dovich International Center for Astrophysics, Bashkir State Pedagogical University, 3A, October Revolution Street, Ufa 450008, RB, Russia}

\author{K.K. Nandi}
\affiliation{Zel'dovich International Center for Astrophysics, Bashkir State Pedagogical University, 3A, October Revolution Street, Ufa 450008, RB, Russia}
\affiliation{High Energy Cosmic Ray Research Center, University of North Bengal, Darjeeling 734 013, WB, India}


\begin{abstract}
A new rotating generalization of the Damour-Solodukhin wormhole (RDSWH), called Kerr-like wormhole, has recently been proposed and investigated by Bueno \textit{et al} for echoes in the gravitational wave signal. We show a novel feature of the RDSWH, viz., that the kinematic properties such as the ISCO\ or marginally stable radius $r_{\textmd{\scriptsize{ms}}}$, efficiency $\epsilon$ and the disk potential $V_{\textmd{\scriptsize{eff}}}$ are \textit{independent} of $\lambda$ (which means they are identical to their KBH counterparts for any given spin). Differences however appear in the emissivity properties for higher values $0.1<\lambda\leq 1$ (say) and for the extreme spin $a_{\star}=0.998$. The kinematic and emissivity are generic properties as variations of the wormhole mass and the rate of accretion within the model preserve these properties. Specifically, the behavior of the luminosity peak is quite opposite to each other for the two objects, which could be useful from the viewpoint of observations. Apart from this, an estimate of the difference $\Delta_{\lambda}$ in the maxima of flux of radiation $F(r)$ shows non-zero values but is too tiny to be observable at present for $\lambda < 10^{-3}$ permitted by the strong lensing bound. The broad conclusion is that RDSWH\ are experimentally indistinguishable from KBH by accretion characteristics.
\end{abstract}

\maketitle


\section{Introduction}

\label{intro}
There is a renewed interest in wormholes after it has been realized that they can mimic post-merger ring-down initial quasi-normal mode (QNM) spectrum of gravitational waves. To our knowledge, this possibility of "wormhole QNM mode" was first put forward by Damour and Solodukhin \cite{Damour:2007}, where resonances were trapped in a double-hump potential associated with what they termed "black hole foil". This is an example of a horizonless wormhole more commonly known as the Damour-Solodukhin wormhole (DSWH). The DSWH differs from the Schwarzschild black hole (SBH) by a dimensionless real deviation parameter $\lambda $ and represents a twice-asymptotically flat regular spacetime connected by a throat. The authors showed that if $\lambda$ is so small that the time scale $\Delta t=2GM\ln \left(\frac{1}{\lambda^{2}}\right) $\ is longer than the observational time scale, the signals emitted by a source falling into a wormhole will contain the usual QNM ringing signature of a black hole, in spite of the absence of a true horizon \cite{Damour:2007}. V\"{o}lkel and Kokkotas \cite{Volkel:2018} have recently shown, using an inverse method, that the knowledge of the observed QNM spectrum can allow one to also accurately construct the double hump P\"{o}schl--Teller potential approximating that of DSWH. Bueno \emph{et al.} \cite{Bueno:2018} suitably redefined the static DSWH and further generalized it the into a Kerr-like wormhole, called here the rotating DSWH or RDSWH for brevity. They studied its gravitational wave echo properties. Gravitational deflection of relativistic massive particles by RDSWH has been recently studied by Jusufi \emph{et al.} \cite{Jusufi:2019}.

QNM ringing by BHs have been well reviewed, see, e.g. Berti \emph{et al.} \cite{Berti:2009} but ringing by WHs still remained an open question. A first step towards answering the question was initiated by Cardoso, Franzin and Pani \cite{Cardoso:2016}, who studied the QNM ringing using a wormhole assembled by means of Visser's cut-and-paste surgery of two copies of Schwarzschild black holes (SBH) at a radius close to the horizon \cite{Visser:1995}. (For future perspectives and new directions of research on ultracompact objects (UCO) including wormholes, see \cite{Cardoso:2017}.) The authors of \cite{Visser:1995} showed that, while the time evolution of the early QNMs accurately mimic those from SBH horizon, the differences (if any) would appear only at later times. This work inspired an investigation in \cite{Konoplya:2016}, where it has been shown that the massless Ellis-Bronnikov wormhole (EBWH) \cite{Ellis:1973,Bronnikov:1973}, made of the minimally coupled exotic scalar field, can also reproduce the black hole QNM spectrum in the eikonal limit (large $\ell$). Ringing by massive EBWH was studied in \cite{Nandi:2017}. These developments prompt a natural inquiry as to whether static DSWH can exhibit mimicking of SBH in other phenomenon as well. A recent work \cite{Nandi:2018} shows that it indeed can, e.g., it can very accurately mimic SBH strong field lensing properties for $\lambda\leq 10^{-3}$. However, this is only an upper bound, which obviously does not rule out values of $\lambda $ \textit{arbitrarily} close to zero. Strong field lensing is an excellent diagnostic for studying the signatures of WHs versus BHs, see, e.g.,\cite{Izmailov:2019}.

There is another important diagnostic, namely, the accretion phenomenon around compact objects that has already been a very active field of research (see, e.g., \cite{Broderick:2007,Torres:2002,Zhang:2017,Lin:2015a,Bambi:2011,Bambi:2012,Bambi:2014,Babichev:2004,Pun:2008a,Pun:2008b,Karimov:2018}). The first comprehensive study of accretion disks using a Newtonian approach was made in \cite{Shakura:1973}. Later a general relativistic model of thin accretion disk was developed in three seminal papers by Novikov and Thorne \cite{Novikov:1973}, Page and Thorne \cite{Page:1974} and Thorne \cite{Thorne:1974} under the assumption that the disk is in a steady-state, that is, the mass accretion rate $\dot{M}$ is constant in time and does not depend of the radius of the disk. The disk is further supposed to be in hydrodynamic and thermodynamic equilibrium, which ensure a black body electromagnetic spectrum and properties of emitted radiation. The thin accretion disk model further assumes that individual particles are moving on Keplerian orbits, but for this to be true the central object is assumed to have weak magnetic field, otherwise the orbits in the inner edge of the disk will be deformed.

Accretion around wormholes has also been an active area of research \cite{Harko:2008,Harko:2009a}, more so because wormholes are special types of objects sourced by materials that violate at least the Null Energy Condition (hence exotic). Research included also other types of objects, e.g., quark, boson and fermion stars, brane-world black holes, gravastars, naked singularities (NS) \cite{Bambi:2009,Bambi:2013a,Bambi:2015,Bambi:2017,Lin:2015b,Bhattacharyya:2001a,Kovacs:2009,Yuan:2004,Guzman:2005,Harko:2009b,Avara:2016,Chen:2011,Kovacs:2010,Danila:2015,Chowdhury:2012,Joshi:2014,Perez:2017}, $f(R)$-modified gravity models of black holes \cite{Perez:2013,Staykov:2016,Ghasemi:2018} and so on. One of the most promising method to distinguish different types of astrophysical objects through their accretion disk properties is the profile analysis of $K\alpha$ iron line \cite{Meyer:2017,Bambi:2013b,Bambi:2013c,Bambi:2016,Jiang:2016}.

In this paper, we shall study the kinematic as well as emissivity properties such as the luminosity spectra, flux of radiation, temperature profile, efficiency of a thin accretion disk around a stellar sized RDSWH using the Page-Thorne model \cite{Shakura:1973,Novikov:1973,Page:1974,Thorne:1974}. As a numerical example, we shall assume the accretion to take place only in the attractive positive mass mouth with mass $15M_{\odot }$ and an accretion rate $\dot{M_{0}}\sim 10^{19}$ gm.sec$^{-1}$. The motive is to analyze how far the accretion profiles are influenced by $\lambda$ and the dimensionless spin parameter $a_{\ast }$.

The paper is organized as follows: In Sec.2, we outline the main formulas relating to the thin accretion disk to be used in the paper. The kinematic accretion formulas for generic spinning spacetime are presented in Sec.3. These are applied to RDSWH in Sec.4. Numerical estimates are made in Sec.5, while Sec.6 concludes the paper. We take units such that $G=c=1$ unless restored and metric signature ($-,+,+,+$).

\section{Thin accretion disk}

\label{sec:2}
We assume geometrically thin accretion disk, which means that the disk height $H$ above the equator is much smaller than the characteristic radius $R$ of the disk, $H\ll R$. The disk is assumed to be in hydrodynamical equilibrium stabilizing its vertical size, with the pressure and vertical entropy gradient being negligible. An efficient cooling mechanism via heat loss by radiation over the disk surface is assumed to be functioning in the disk, which prevents the disk from collecting the heat generated by stresses and dynamical friction. The thin disk has an inner edge defined by the marginally stable circular radius $r_{\textmd{\scriptsize{ms}}}$, while the orbits at higher radii are Keplerian. In the steady-state approximation, the mass accretion rate $\dot{M}_{0}$ is assumed to be a constant and the physical quantities describing the accreting matter are averaged over a characteristic time scale $\Delta t$, over the azimuthal angle $\Delta \phi = 2\pi $ for a total period of the orbits, and over the height $H$ \cite{Novikov:1973,Page:1974}.

The orbiting particles with the four-velocity $u^{\mu}$ form a disk of an averaged surface density $\Sigma$, where the rest mass density $\rho_{0}$, the energy flow vector $q^{\mu}$ and the stress tensor $t^{\mu\nu}$ are measured in the averaged rest-frame. Then
\begin{equation}
\Sigma (r)=\int_{-H}^{H}\langle \rho_{0}\rangle dz,
\end{equation}%
where $\langle \rho _{0}\rangle $ rest mass density averaged over $\Delta t$ and $2\pi$ and the torque density
\begin{equation}
W_{\phi }{}^{r}=\int_{-H}^{H}\langle t_{\phi }{}^{r}\rangle dz,
\end{equation}%
with the component $\langle t_{\phi }^{r}\rangle $ averaged over $\Delta t$ and $2\pi$. The time and orbital average of the energy flow vector $q^{\mu}$ gives the radiation flux ${\mathcal{F}}(r)$ over the disk surface as
\begin{equation}
{\mathcal{F}}(r)=\langle q^{z}\rangle .
\end{equation}

For the stress energy tensor $T_{\nu}^{\mu}$ of the disk, the energy and angular momentum four-vectors are defined by $-E^{\mu}\equiv T_{\nu}^{\mu}(\partial/\partial t)^{\nu}$ and $J^{\mu}\equiv T_{\nu}^{\mu}(\partial/\partial\phi)^{\nu}$ respectively. The structure equations of the thin disk can be derived by integrating the conservation laws of the rest mass, of the energy, and of the angular momentum \cite{Novikov:1973,Page:1974}. From the rest mass conservation, $\nabla_{\mu}(\rho_{0}u^{\mu})=0$, it follows that the average rate of the accretion is independent of the disk radius,
\begin{equation}
\dot{M_{0}}\equiv -2\pi r\Sigma u^{r}=\mathrm{constant}.
\end{equation}

In the steady-state approximation, specific energy $\widetilde{E}$ and specific angular momentum $\widetilde{L}$ of accreting particles have depend only on the radius of the orbits. Defining black hole rotational velocity $\Omega =d\phi /dt$, the energy conservation law $\nabla_{\mu}E^{\mu}=0$ yields the integral
\begin{equation}
\lbrack \dot{M}_{0}\widetilde{E}-2\pi r\Omega W_{\phi }{}^{r}]_{,r}=4\pi r{\mathcal{F}}\widetilde{E}.
\end{equation}%
This is a balance equation, which states that the energy transported by the rest mass flow, $\dot{M}_{0}\widetilde{E}$, and the energy transported by the torque in the disk, $2\pi r\Omega W_{\phi}^{r}$, is balanced by the energy radiated away from the surface of the disk, $4\pi r{\mathcal{F}}\widetilde{E}$.

The angular momentum conservation law, $\nabla_{\mu}J^{\mu}=0$, states the balance of three forms of angular momentum transport, viz.,
\begin{equation}
\lbrack \dot{M}_{0}\widetilde{L}-2\pi rW_{\phi }{}^{r}]_{,r}=4\pi r{\mathcal{F}}\widetilde{L}\;.
\end{equation}%
By eliminating $W_{\phi}^{r}$ from Eqs. (5) and (6), and applying the energy-angular momentum relation for circular geodesic orbits in the form $\widetilde{E}_{,r}=\Omega \widetilde{L}_{,r}$, the flux $F$ of the radiant energy, or power, over the disk can be expressed as \cite{Novikov:1973,Page:1974},
\begin{equation}
F\left(r\right) =-\frac{\dot{M_{0}}}{4\pi\sqrt{-g}}\frac{\Omega _{,r}}{\left( \widetilde{E}-\Omega \widetilde{L}\right) ^{2}}\int_{r_{\textmd{\scriptsize{ms}}}}^{r}\left( \widetilde{E}-\Omega \widetilde{L}\right) \widetilde{L}_{,r}dr.
\end{equation}%
The disk is supposed to be in thermodynamical equilibrium, as explained, so the radiation flux emitted by the disk surface will follow Stefan-Boltzmann law:%
\begin{equation}
F\left(r\right) =\sigma T^{4}\left( r\right) ,
\end{equation}%
where $\sigma $ is the Stefan-Boltzmann constant. The observed luminosity $L\left(\nu\right)$ has a redshifted black body spectrum \cite{Torres:2002}
\begin{equation}
L_{\nu}=4\pi \textit{d}^{2}I(\nu)=\frac{8\pi h\cos{i}}{c^{2}}\int_{r_{\textmd{\scriptsize{in}}}}^{r_{\textmd{\scriptsize{out}}}}\int_{0}^{2\pi }\frac{\nu_{e}^{3}rdrd\varphi}{\textmd{exp}\left[\frac{h\nu_{e}}{\kappa_{B}T}\right]
-1},
\end{equation}%
where $i$ is the disk inclination angle to the vertical, \textit{d} is the distance between the observer and the center of the disk, $r_{\textmd{\scriptsize{in}}}$ and $r_{\textmd{\scriptsize{out}}}$ are the inner and outer radii of the disc, $h$ is the Planck constant, $\nu_{e}$ is the emission frequency, $I(\nu)$ is the Planck distribution, and $k_{B}$ is the Boltzmann constant. The observed photons are redshifted and received frequency $\nu$ is related to the emitted ones by $\nu_{e}=(1+z)\nu$. The redshift factor $(1+z)$ has the form:
\begin{equation}
1+z=\frac{1+\Omega r \sin \phi \sin i}{\sqrt{ -g_{tt}-2\Omega g_{t\phi} -
\Omega^2 g_{\phi\phi}}},
\end{equation}%
where the light bending effect is neglected \cite{Luminet:1979,Bhattacharyya:2001b}.

Another important characteristic of the thin accretion disk is its efficiency $\epsilon$, which quantifies the ability with which the central body converts the accreting mass into radiation. The efficiency is measured at infinity and it is defined as the ratio of two rates: the rate of energy of the photons emitted from the disk surface and the rate with which the mass-energy is transported to the central body. If all photons reach infinity, the Page-Thorne accretion efficiency is given by the specific energy of the accreting particles measured at the marginally stable orbit \cite{Page:1974}:
\begin{equation}
\epsilon =1-\widetilde{E}\left(r_{\textmd{\scriptsize{ms}}}\right),
\end{equation}%
As the definition indicates, $\epsilon$ should be non-negative.

\section{Generic rotating spacetime}

\label{sec:3}
The accretion disk is formed by particles moving in circular orbits around a compact object, with the geodesics determined by the space-time geometry around the object, be it a WH, BH or NS. For a rotating geometry the metric is generically given by
\begin{equation}
ds^{2} = -g_{tt}\,dt^{2} + 2g_{t\phi}\,dtd\phi + g_{rr}\,dr^{2} + g_{\theta\theta}\,d\theta^{2} + g_{\phi\phi}\,d\phi^{2}\,.
\end{equation}%
At and around the equator, i.e., when $\left\vert \theta -\pi /2\right\vert \ll 1,$we assume, with Harko \emph{et al} \cite{Harko:2009b}, that the metric functions $g_{tt}$, $g_{t\phi}$, $g_{rr}$, $g_{\theta\theta}$ and $g_{\phi\phi}$ depend only on the radial coordinate $r$. The angular velocity $\Omega $, of the specific energy $\widetilde{E}$, and of the specific angular momentum $\widetilde{L}$ of accreting particles in the above geometry are given by
\begin{eqnarray}
\frac{dt}{d\tau } &=&\frac{\widetilde{E}g_{\phi \phi }+\widetilde{L}g_{t\phi}}{g_{t\phi }^{2}+g_{tt}g_{\phi \phi }}\,, \\
\frac{d\phi }{d\tau } &=&-\frac{\widetilde{E}g_{t\phi}-\widetilde{L}g_{tt}}{g_{t\phi }^{2}+g_{tt}g_{\phi \phi }}\,, \\
g_{rr}\left( \frac{dr}{d\tau }\right) ^{2} &=&-1+\frac{\widetilde{E}^{2}g_{\phi \phi }+2\widetilde{E}\widetilde{L}g_{t\phi }-\widetilde{L}^{2}g_{tt}}{g_{t\phi }^{2}+g_{tt}g_{\phi \phi }}\,.
\end{eqnarray}

One may hence define an effective potential term defined as
\begin{equation}
V_{\textmd{\scriptsize{eff}}}(r) = -1 + \frac{\widetilde{E}^{2}g_{\phi\phi} + 2\widetilde{E}\widetilde{L}g_{t\phi} - \widetilde{L}^{2}g_{tt}}{g_{t\phi}^{2} + g_{tt}g_{\phi\phi}}\,.
\end{equation}%
Existence of circular orbits at any arbitrary radius $r$ in the equatorial plane demands that $V_{\textmd{\scriptsize{eff}}}(r)=0$ and $dV_{\textmd{\scriptsize{eff}}}/dr=0$. These conditions allow us to write the kinematic parameters as
\begin{eqnarray}
\widetilde{E} &=&\frac{g_{tt}-g_{t\phi}\Omega}{\sqrt{g_{tt}-2g_{t\phi}\Omega - g_{\phi\phi}\Omega^{2}}}\,, \\
\widetilde{L} &=&\frac{g_{t\phi}+g_{\phi\phi}\Omega}{\sqrt{g_{tt} - 2g_{t\phi}\Omega - g_{\phi\phi}\Omega^{2}}}\,, \\
\Omega &=&\frac{d\phi}{dt}=\frac{-g_{t\phi ,r}+\sqrt{(g_{t\phi,r})^{2}+g_{tt,r}g_{\phi \phi ,r}}}{g_{\phi \phi ,r}}\,.
\end{eqnarray}%
where, throughout the paper, $X_{,r}\equiv dX/dr$. Stability of orbits depends on the signs of $d^{2}V_{\textmd{\scriptsize{eff}}}/dr^{2}$, while the condition $d^{2}V_{\textmd{\scriptsize{eff}}}/dr^{2}=0$ gives the inflection point or \textit{marginally stable} (ms) orbit or innermost stable circular orbit (ISCO) at $r=r_{\textmd{\scriptsize{ms}}}$. The thin disk is assumed to have an inner edge defined by the marginally stable circular radius $r_{\textmd{\scriptsize{ms}}}$, while the orbits at higher radii than $r=r_{\textmd{\scriptsize{ms}}}$ are Keplerian. Note that marginally stable circular orbits play a crucial role when an accreting BH is surrounded, e.g., by quintessence matter \cite{Izmailov:2019,Hussain:2016}. The generic Eqs.(7-19) are valid for any rotating spacetime that will be explicitly calculated in what follows.

\section{Accretion disk properties of RDSWH}

We start with a Kerr-like wormhole spacetime, recently considered by Bueno \emph{et al.} \cite{Bueno:2018} that generalized the static Damour-Solodukhin wormhole \cite{Damour:2007}, which we call here RDSWH. The spacetime metric in Boyer-Lindquist coordinates is given by
\begin{eqnarray}
ds^{2} &=& -\left(1-\frac{2Mr}{\Sigma}\right) dt^{2} - \frac{4Mar\sin^{2}{\theta}}{\Sigma}dtd\phi + \frac{\Sigma}{\hat{\Delta}}dr^{2} \nonumber \\
&&+ \Sigma d\theta^{2} + \left(r^{2}+a^{2}+\frac{2Ma^{2}r\sin^{2}{\theta}}{\Sigma}\right)\sin^{2}{\theta}d\phi^{2},
\end{eqnarray}
where $\Sigma$ and $\hat{\Delta}$ are expressed by
\begin{equation}
\Sigma\equiv r^{2}+a^{2}\cos^{2}{\theta},\quad \hat{\Delta}\equiv r^{2}-2M(1+\lambda^{2})r+a^{2}.
\end{equation}

\begin{figure}
  \begin{center}
     \includegraphics[type=pdf,ext=.pdf,read=.pdf,width=8.5cm]{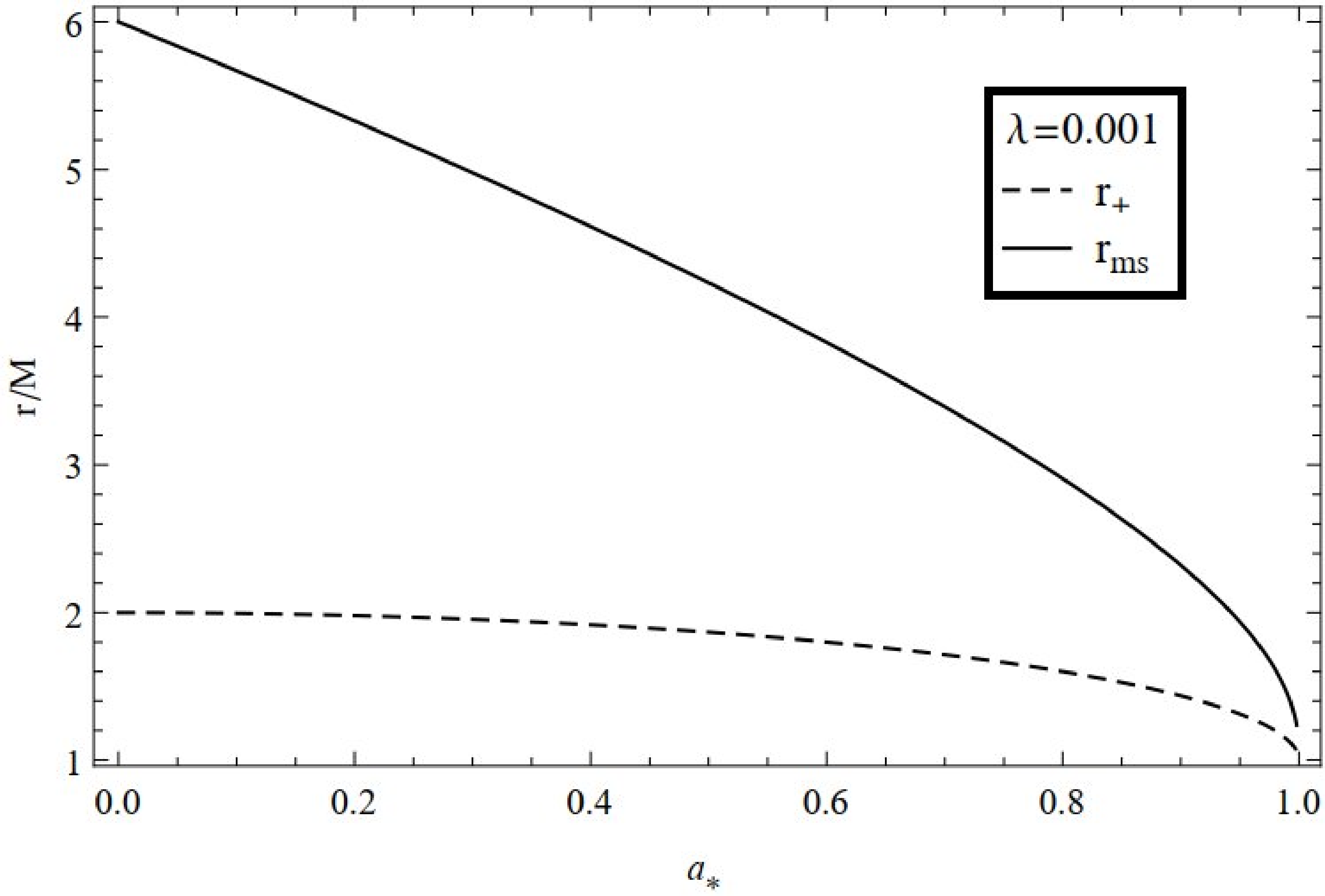}
  \end{center}
  \caption{The value of the radial coordinate for the throat $r_{+}$ (dashed line) and marginally stable $r_{ms}$ (solid line) orbits as functions of the spin parameter $a_{\star}$ for the $\lambda = 10^{-3}$.}
\end{figure}

\begin{figure*}
\begin{center}
\includegraphics[type=pdf,ext=.pdf,read=.pdf,width=8.5cm]{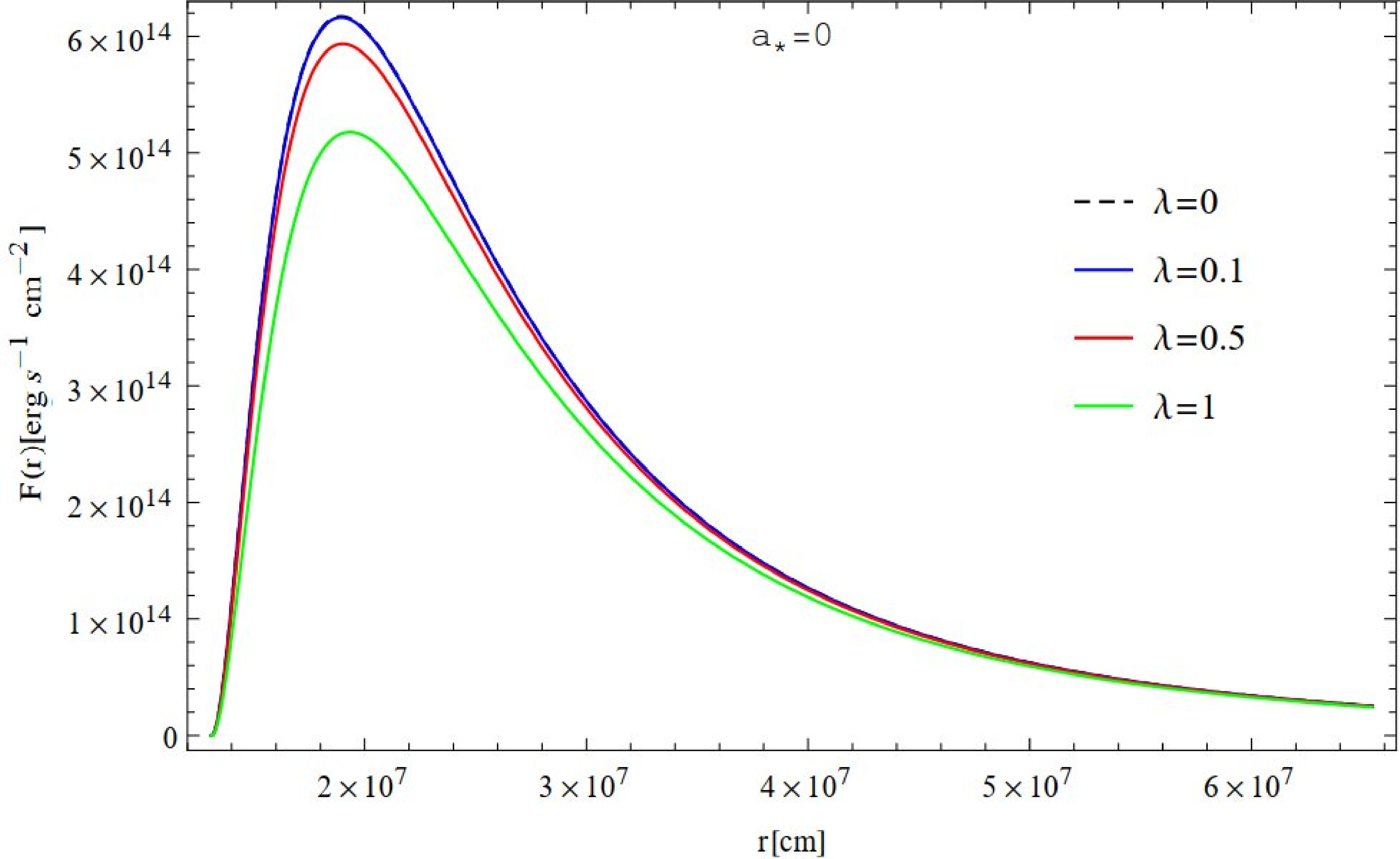}
\hspace{0.3cm}
\includegraphics[type=pdf,ext=.pdf,read=.pdf,width=8.5cm]{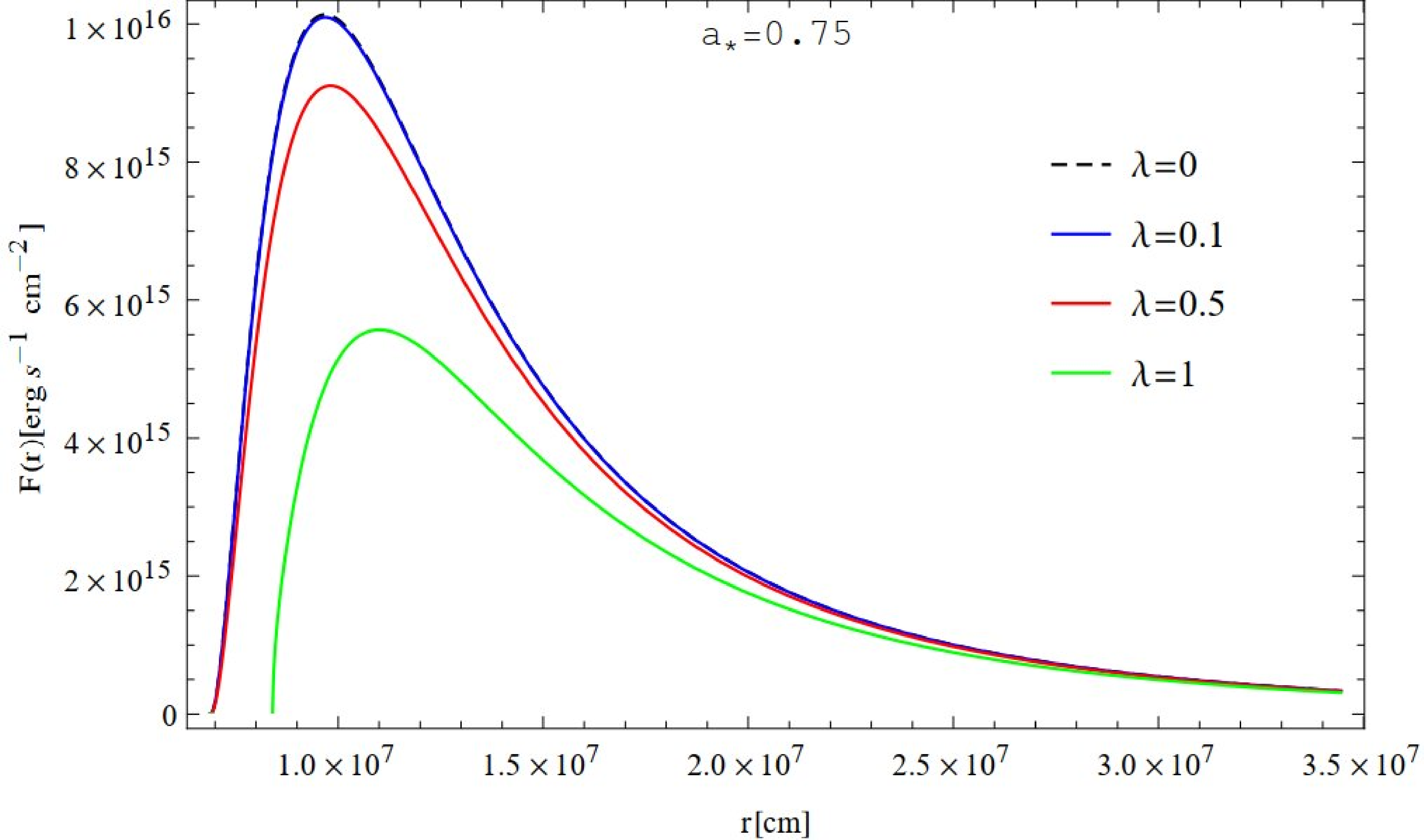} \\
\vspace{0.5cm}
\includegraphics[type=pdf,ext=.pdf,read=.pdf,width=8.5cm]{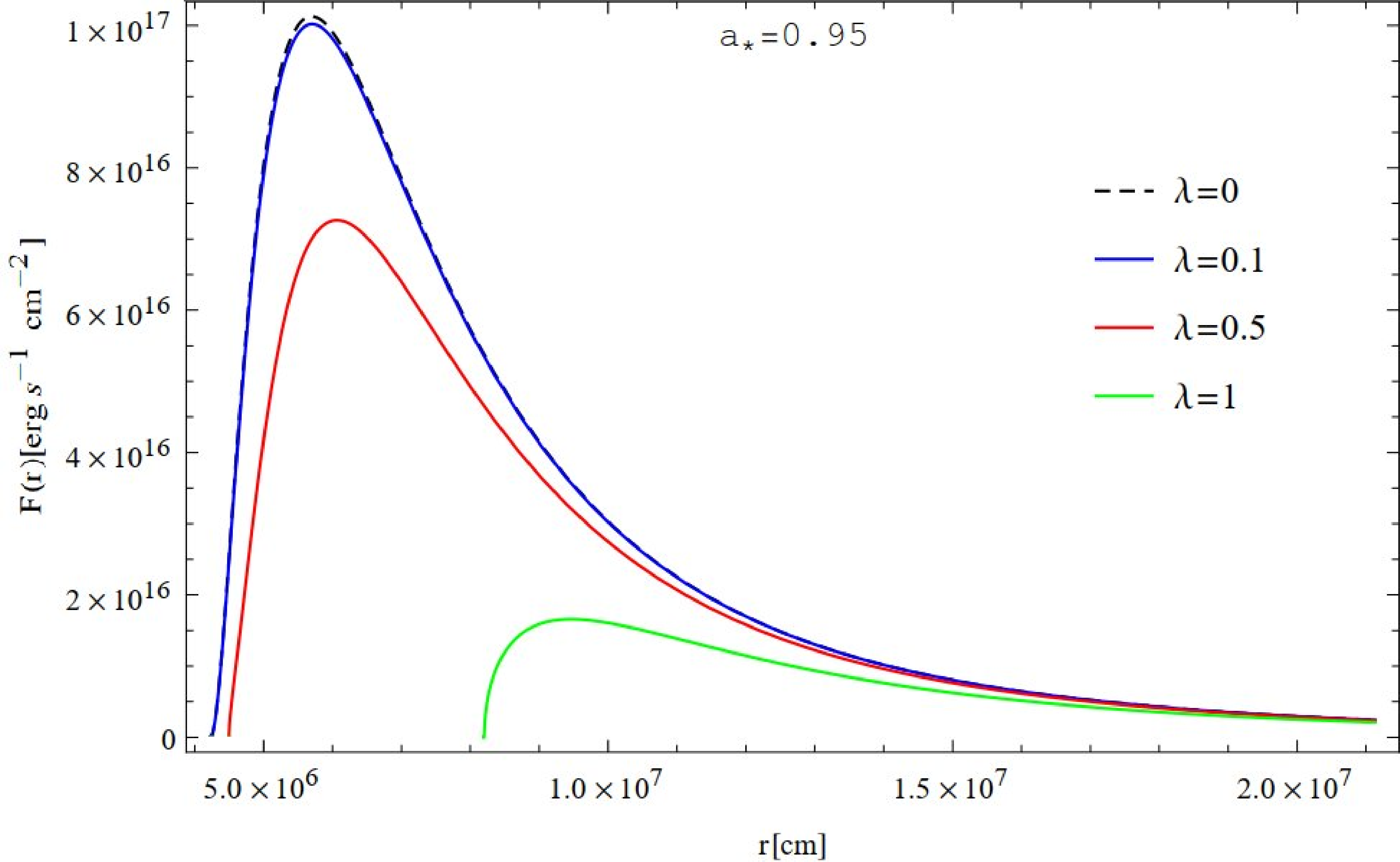}
\hspace{0.3cm}
\includegraphics[type=pdf,ext=.pdf,read=.pdf,width=8.5cm]{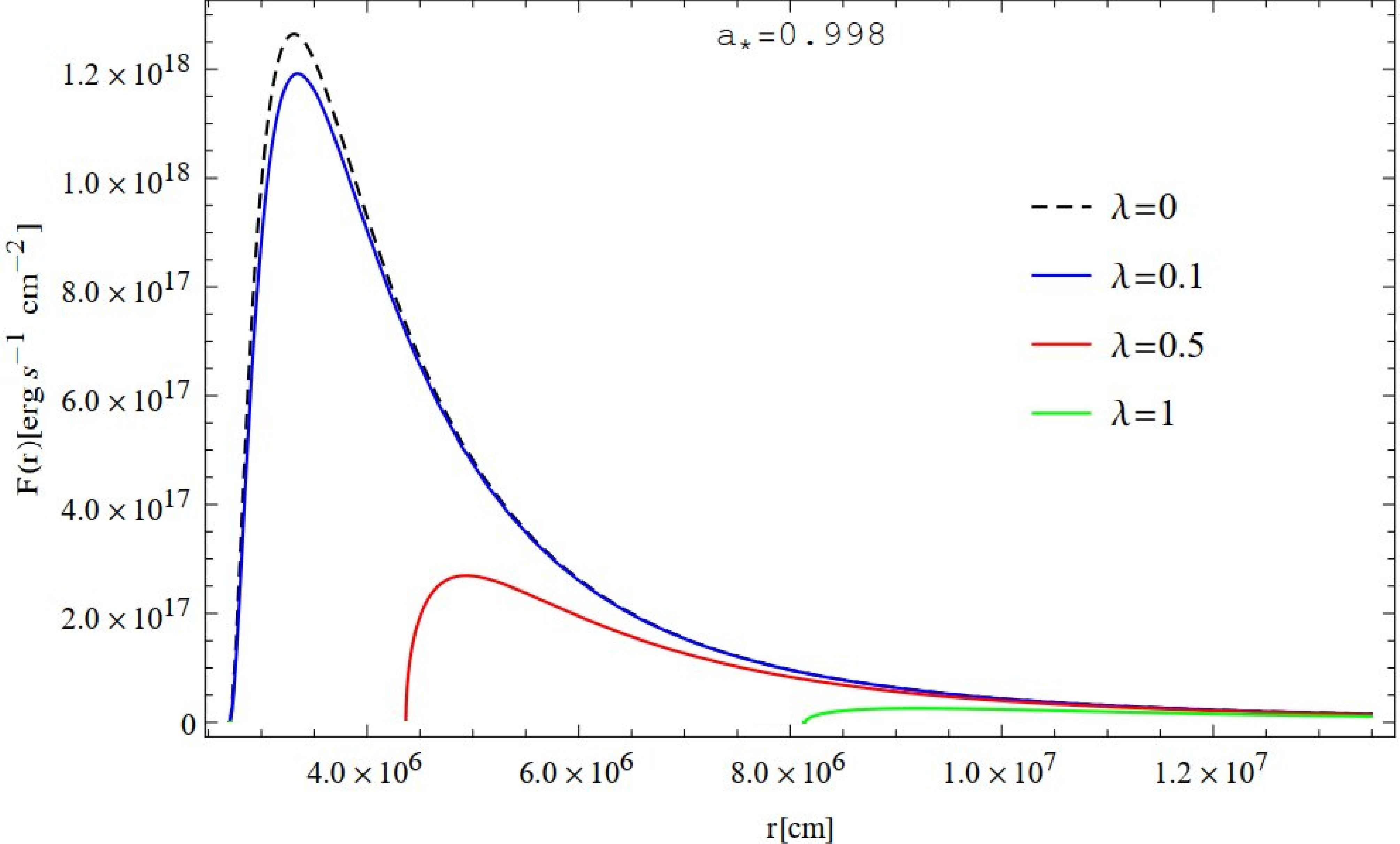}
\end{center}
\caption{The time averaged flux $F(r)$ as a function of the radial coordinate $r$ (in cm) radiated by the disk for a RDSWH plotted for different values of $\protect\lambda$ and compared with the Kerr BH.}
\label{Flux}
\end{figure*}

\begin{figure*}
\begin{center}
\includegraphics[type=pdf,ext=.pdf,read=.pdf,width=8.5cm]{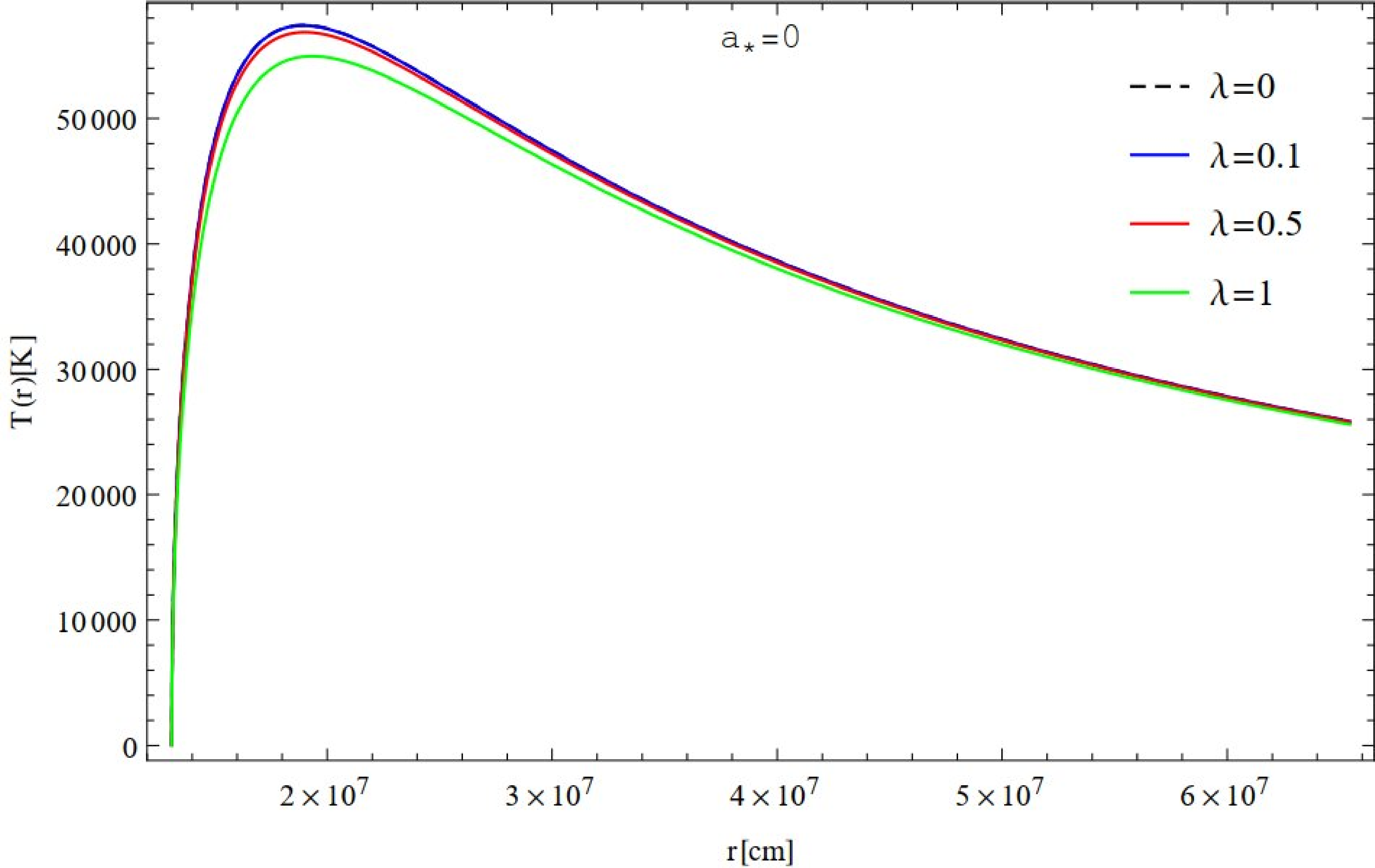}
\hspace{0.3cm}
\includegraphics[type=pdf,ext=.pdf,read=.pdf,width=8.5cm]{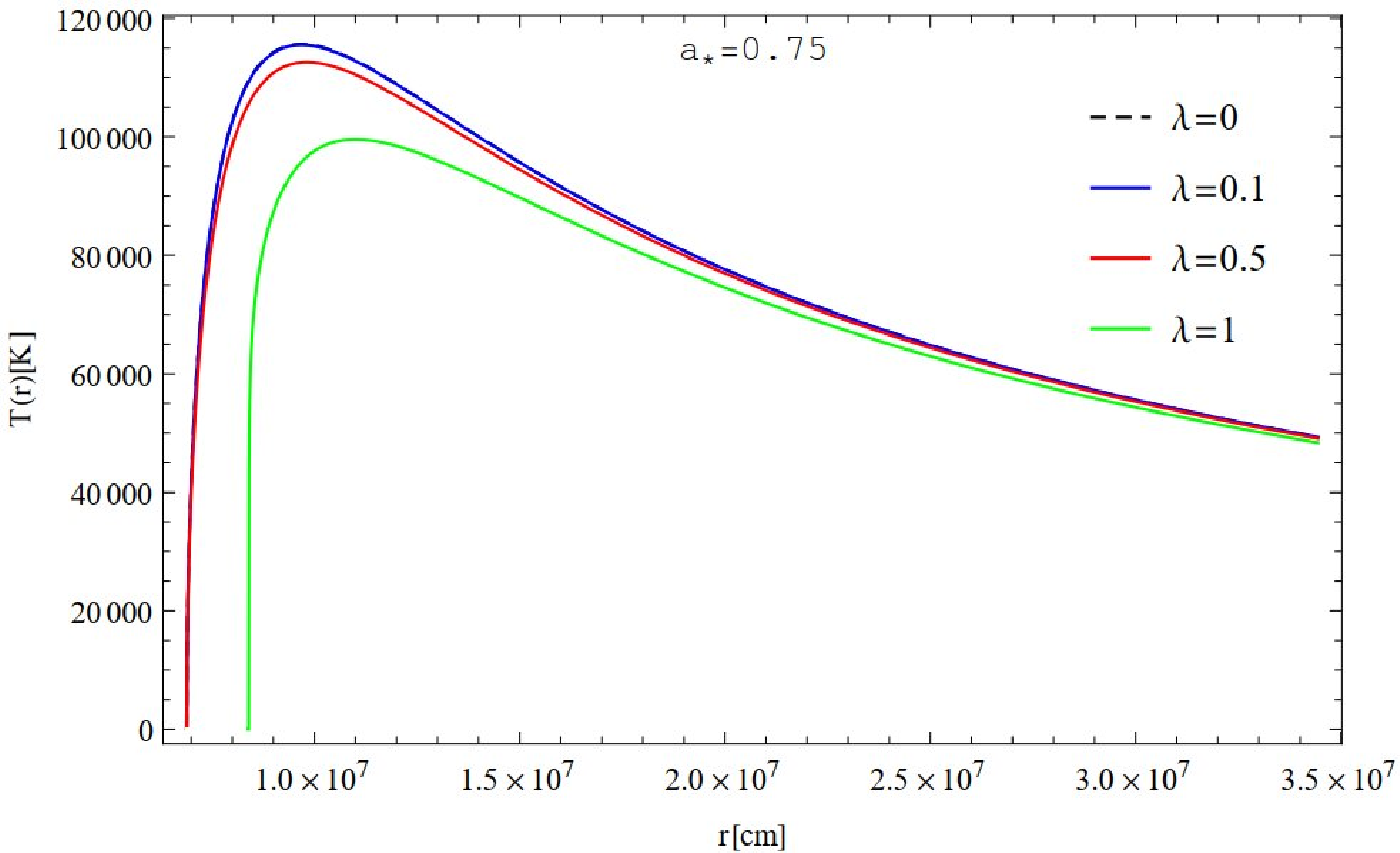} \\
\vspace{0.5cm}
\includegraphics[type=pdf,ext=.pdf,read=.pdf,width=8.5cm]{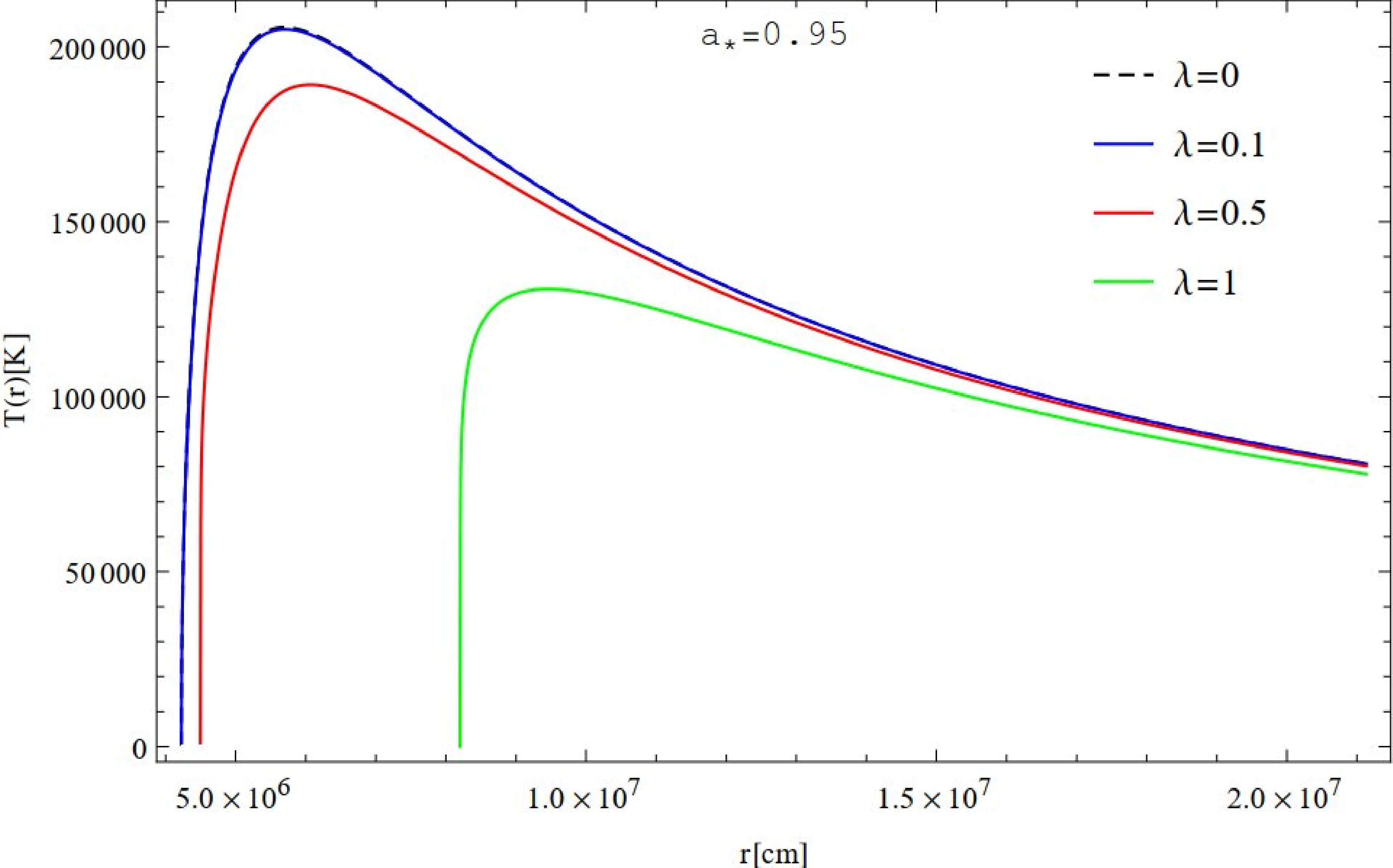}
\hspace{0.3cm}
\includegraphics[type=pdf,ext=.pdf,read=.pdf,width=8.5cm]{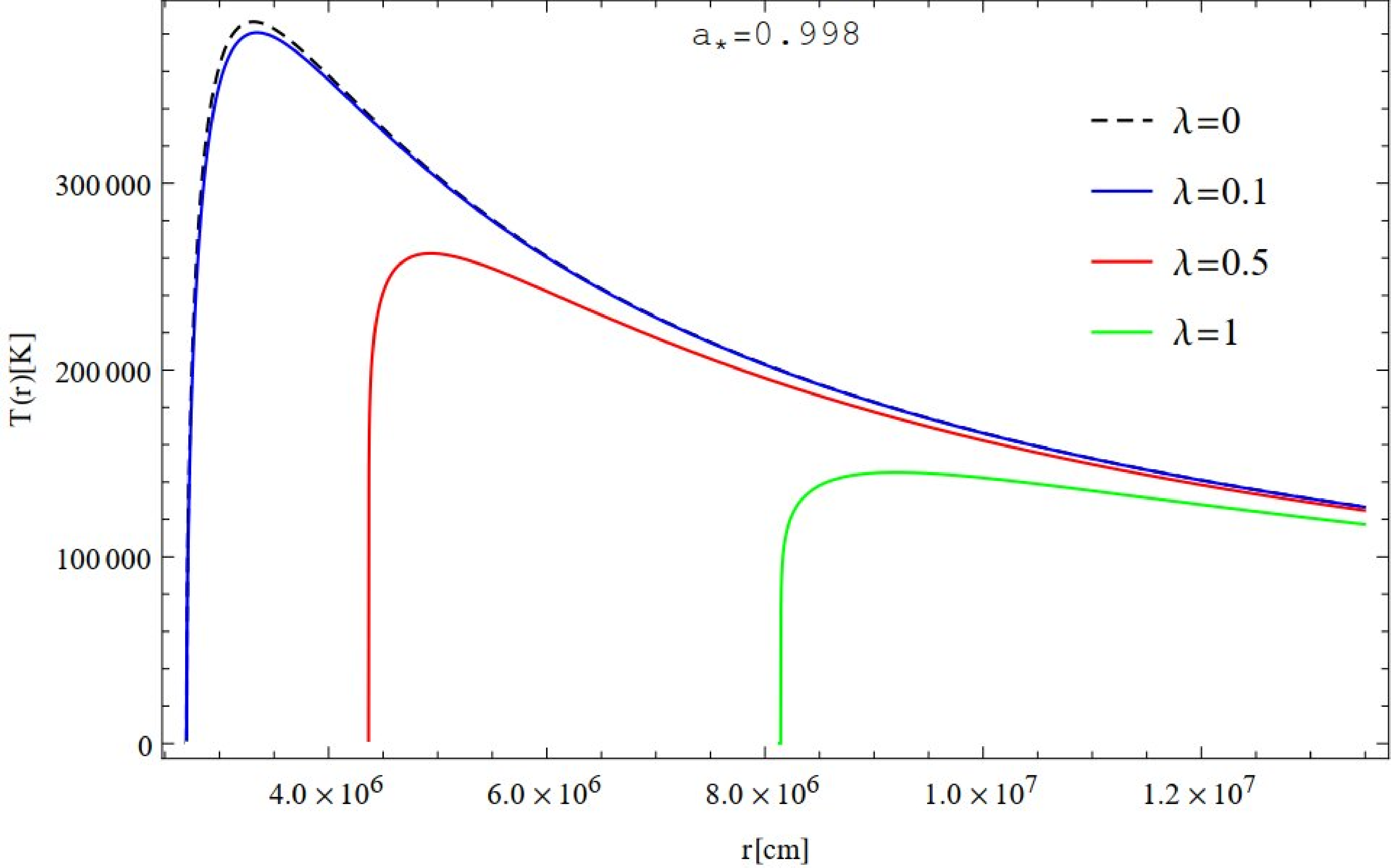}
\end{center}
\caption{Temperature distribution $T(r)$ as a function of the radial coordinate $r$ (in cm) of the accretion disk for a RDSWH plotted for different values of $\protect\lambda $ and compared with the Kerr BH.}
\label{Temp}
\end{figure*}

\begin{figure*}
\begin{center}
\includegraphics[type=pdf,ext=.pdf,read=.pdf,width=8.5cm]{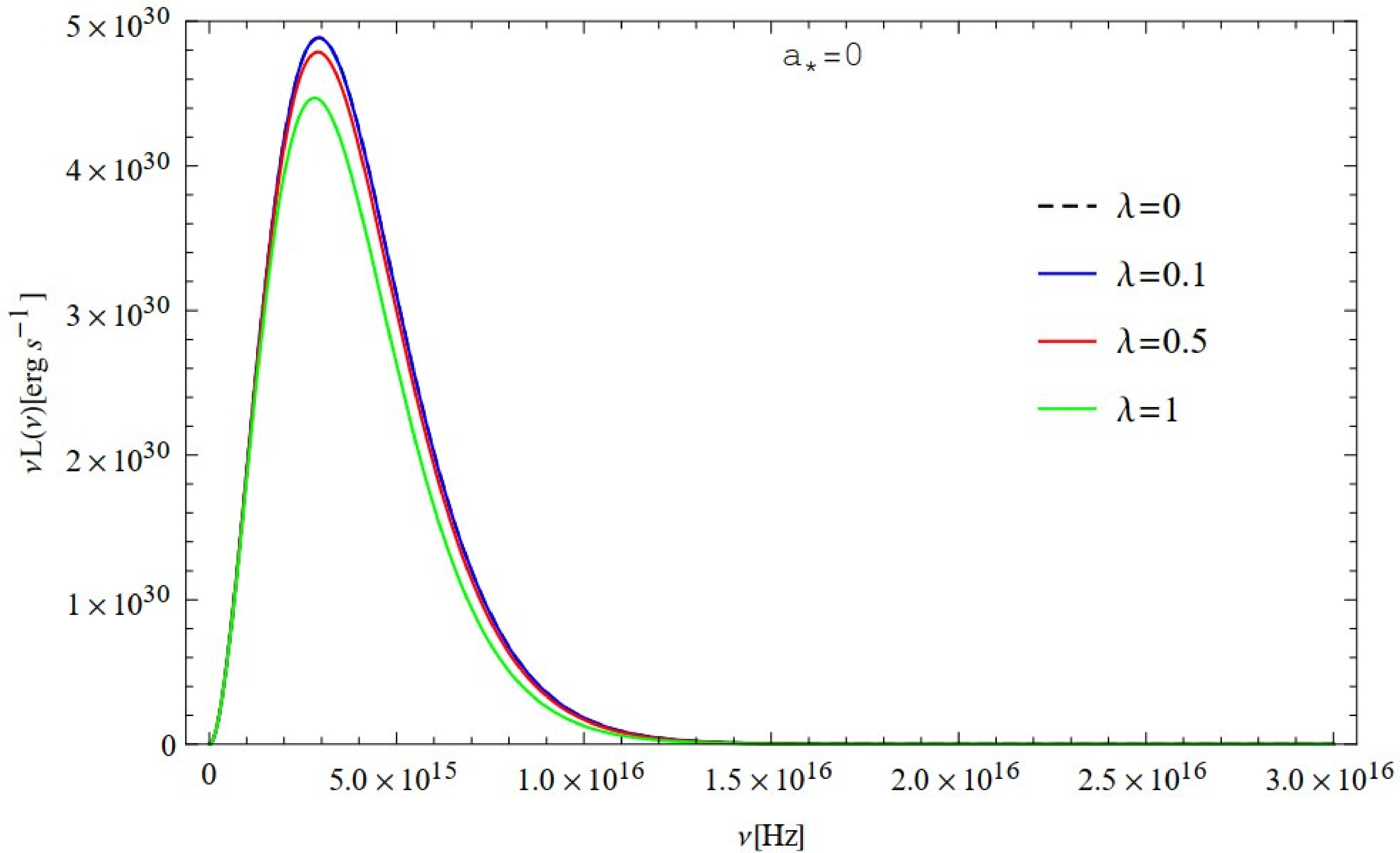}
\hspace{0.3cm}
\includegraphics[type=pdf,ext=.pdf,read=.pdf,width=8.5cm]{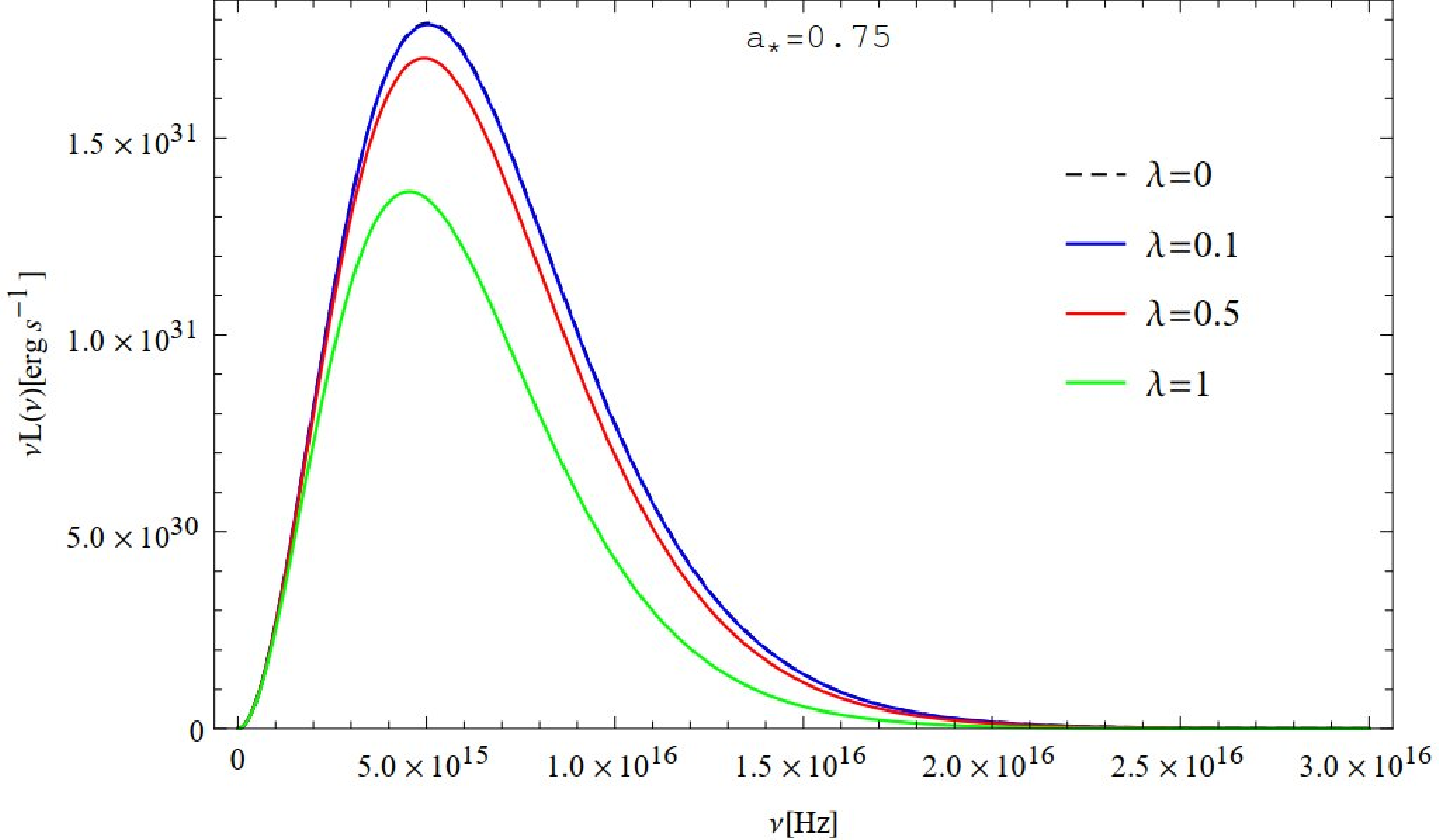} \\
\vspace{0.5cm}
\includegraphics[type=pdf,ext=.pdf,read=.pdf,width=8.5cm]{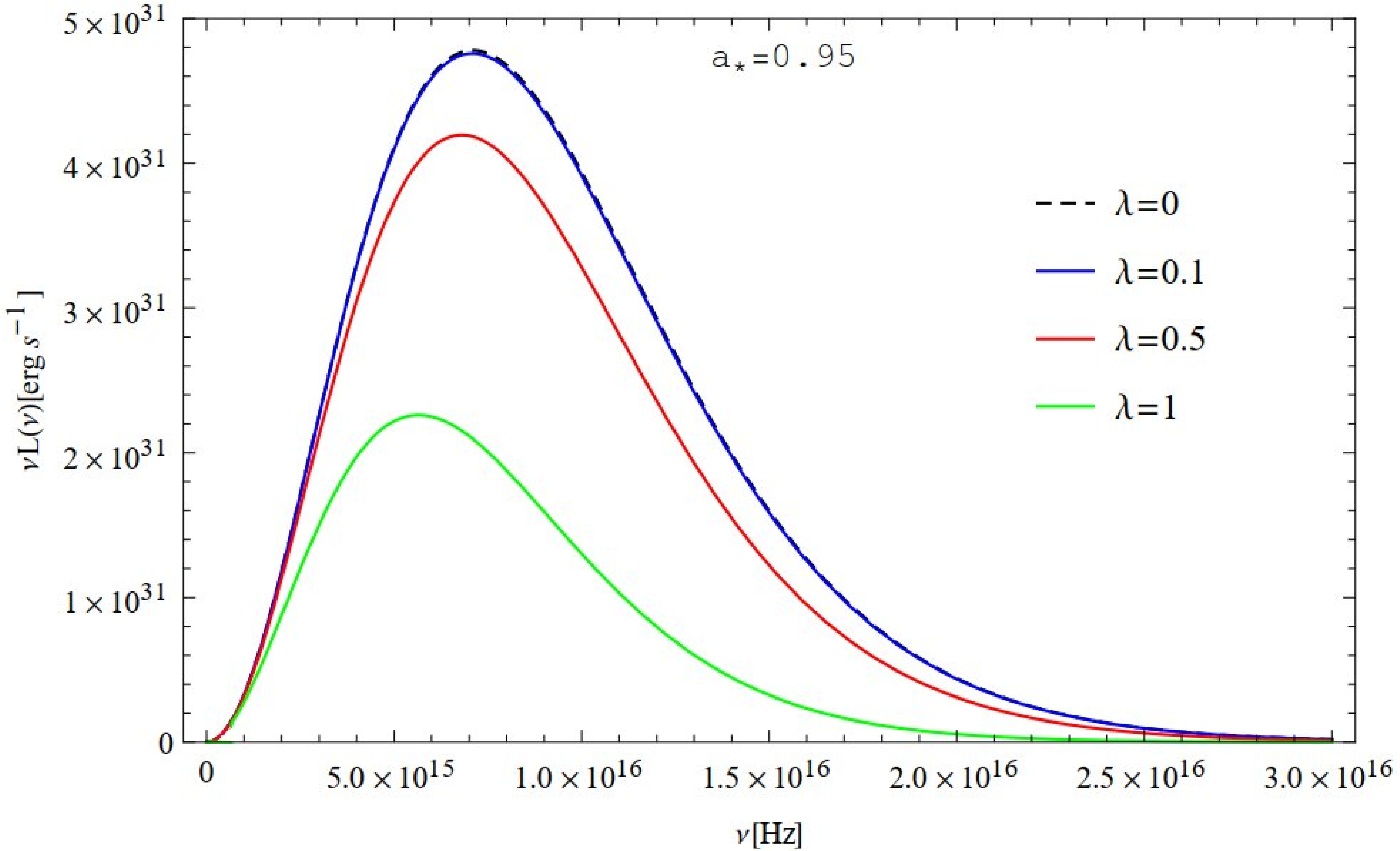}
\hspace{0.3cm}
\includegraphics[type=pdf,ext=.pdf,read=.pdf,width=8.5cm]{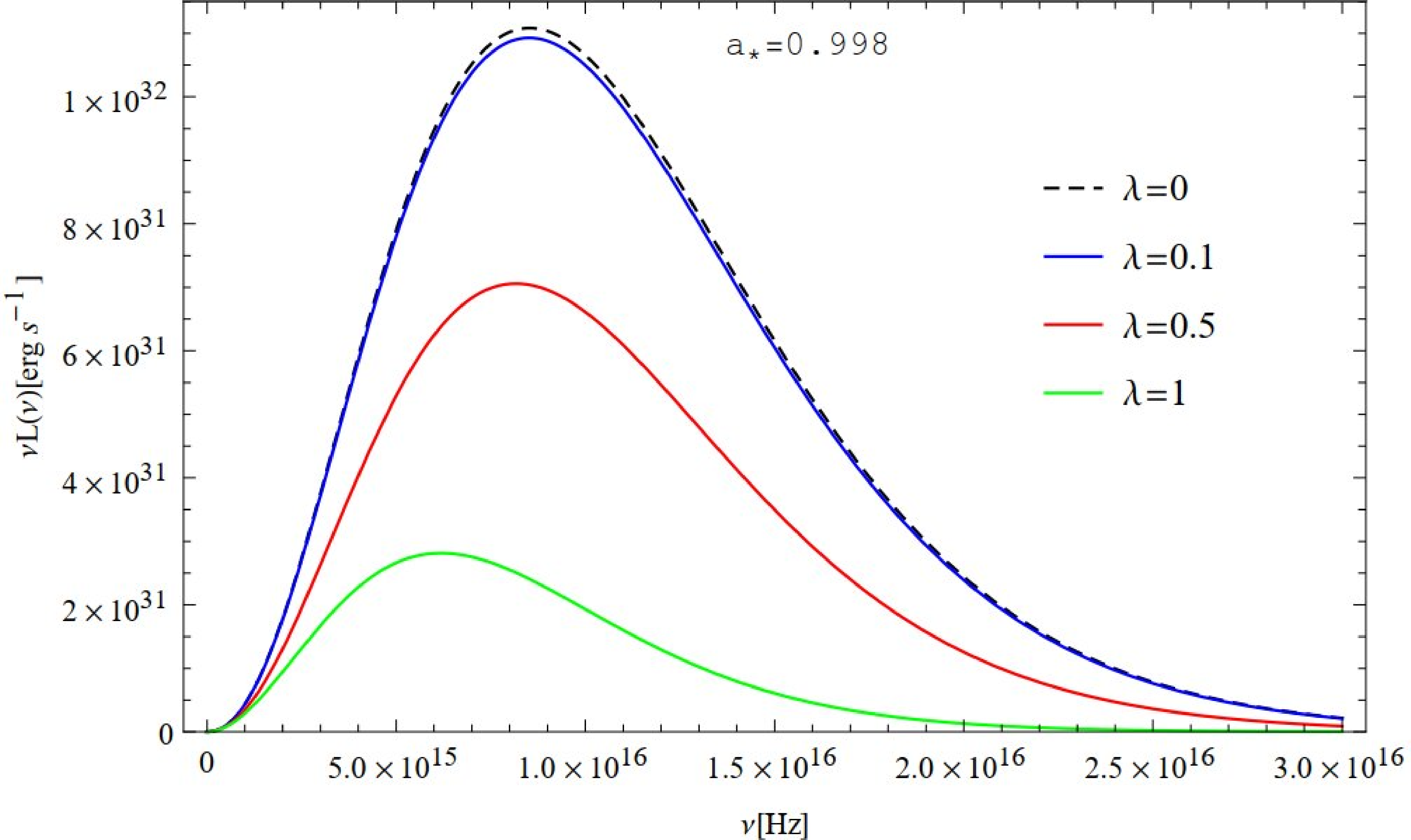}
\end{center}
\caption{The emission spectra $\protect\nu L(\protect\nu )$ of the accretion disk for a RDSWH plotted for different values of $\protect\lambda$ and compared with the Kerr BH.}
\label{Lum}
\end{figure*}

It contains a family of parameters where $a$ and $M$ corresponds to the spin and the mass of wormhole, and $\lambda^{2}$ is the deformation parameter\footnote{We clarify that $M$ here is the Keplerian mass as considered by Bueno \emph{et al.} \cite{Bueno:2018}, viz., $g_{tt}\sim 1-\frac{2M}{r}$ in their Eq.(36) and explicitly so stated after their RDSWH metric (46). Keplerian mass is the one measured by the orbiting particles in the accretion disk or the one responsible for the echoes \cite{Bueno:2018}, both being strong field phenomena. On the other hand, the ADM mass is measured by an asymptotic static observer and is given by \cite{Amir:2019}: $M_{ADM}=M\left( 1+\lambda ^{2}\right) $. While the ADM mass is not relevant for studies in accretion or echoes, the existence of two distinct masses here reveal that the RDSWH possesses characteristics of Machian Brans-Dicke solutions, where the Keplerian mass is distinct from the ADM mass. It would thus be rewarding to relate $\lambda$ to some kind of scalar field and develop a scalar-tensor theory for which RDSWH could be an exact solution. We thank the anonymous reviewer whose query has led us to think of this interesting possibility to be pursued in the future.}. A non vanishing $\lambda^{2}$ differs this metric from Kerr metric, we can recover the Kerr metric when $\lambda^{2}=0$. Although the throat of the wormhole can be easily obtained by equating the $\hat{\Delta}$ to zero,
\begin{equation}
r_{+} = (1+\lambda^{2})M+\sqrt{M^{2}(1+\lambda^{2})^{2}-a^{2}},
\end{equation}%
which represents a special region that connects two different asymptotically flat regions. For any values of $\lambda\neq 0$ the metric is regular everywhere and with small values of $\lambda^{2}\sim 0$, it is practically indistinguishable from a Kerr BH.

Special feature of RDSWH metric is that kinematic properties of accretion disk coincide with those of Kerr BH. As an example let us concider the effective potential, given by Eq.(16), which determines the geodesic motion of the test particles in the equatorial plane of RDSWH solution (20)-(21). This is given by
\begin{equation}
V_{\textmd{\scriptsize{eff}}} = -1 + \frac{2M(\widetilde{L}-a\widetilde{E})^{2}+r(a\widetilde{E}-\widetilde{L})(a\widetilde{E}+\widetilde{L})+\widetilde{E}^{2}r^{3}}{r\left\{ a^{2}+r(r-2M)\right\} }.
\end{equation}%
Here we see that expression for effective potential of RDSWH solution does \textit{not} depend on $\lambda$ and therefore coincides with expression for effective potential of Kerr BH and so the ISCO radius or $r_{\textmd{\scriptsize{ms}}}$ is the same as that of Kerr BH.

The remaining expressions determining kinematic properties of RDSWH are specific energy, specific angular momentum, angular velocity and radius of the marginally stable orbit which follow from Eqs.(17)-(19). They also coincide with those of Kerr BH:
\begin{eqnarray}
\widetilde{E} &=&\frac{r^{2}(r-2M)+aM(2\sqrt{Mr}-a)}{r\sqrt{r^{3}(r-3M)+6a(Mr)^{3/2}-3a^{2}M(r+M)+2a^{3}\sqrt{Mr}}}, \\
\widetilde{L} &=&\frac{\sqrt{Mr^{7}}-3aMr^{2}+a^{2}\sqrt{Mr}(r+2M)-a^{3}M}{r\sqrt{r^{3}(r-3M)+6a(Mr)^{3/2}-3a^{2}M(r+M)+2a^{3}\sqrt{Mr}}}, \\
\Omega  &=&\frac{\sqrt{Mr^{3}}-aM}{r^{3}-a^{2}M}, \\
r_{\textmd{\scriptsize{ms}}} &=&3M+\sqrt{3M^{2}+a^{2}+P}-\frac{1}{2}\left[72M^{2}-8(6M^{2}-a^{2})\right. \nonumber \\
&&\left.-4P+64a^{2}M(3M^{2}+a^{2}+P)^{-\frac{1}{2}}\right]^{1/2},
\end{eqnarray}%
where
\begin{eqnarray}
P &=&\frac{9M^{4}-10a^{2}M^{2}+a^{4}}{K^{\frac{1}{3}}}+K^{\frac{1}{3}},
\nonumber \\
K &=&27M^{6}-45a^{2}M^{4}-8a^{3}M^{3}+17a^{4}M^{2}+8a^{5}M+a^{6}.  \nonumber
\end{eqnarray}%
The reason of coincidence of kinematic properties is that parameter $\lambda$ appears only in the $g_{rr}$ component of metric which is not used in Eqs.(16)-(19).

\begin{table*}[!ht]
\caption{The maximum values of the time averaged radiation flux $F(r)$, temperature distribution $T(r)$ with corresponding critical radii and the emission spectra with corresponding critical frequencies for RDSWH. The critical values of radius $r_{\textmd{\scriptsize{crit}}}$ and of frequency $\nu_{\textmd{\scriptsize{crit}}}$ where the corresponding maxima occur is shown in the columns 5 and 7 respectively.}
\centering
\begin{tabular}{|c|c|c|c|c|c|c|}
\hline
$a_{\star}$ & $\lambda$ & $F_{\textmd{\scriptsize{max}}}(r)$ & $T_{\textmd{\scriptsize{max}}}(r)$ & $r_{\textmd{\scriptsize{crit}}}$ & $\nu L(\nu)_{\textmd{\scriptsize{max}}}$ & $\nu_{\textmd{\scriptsize{crit}}}$ \\
&  & [erg s$^{-1}$ cm$^{-2}$]$\times 10^{14}$ & [K] & [cm]$\times 10^{7}$ & [erg]$\times 10^{30}$ & [Hz]$\times 10^{15}$ \\
\hline
$0$ & $0$ & $6.170$ & $57435$ & $1.893$ & $4.888$ & $2.928$ \\
& $0.1$ & $6.161$ & $57414$ & $1.893$ & $4.883$ & $2.927$ \\
& $0.5$ & $5.936$ & $56881$ & $1.901$ & $4.788$ & $2.902$ \\
& $1$ & $5.180$ & $54976$ & $1.934$ & $4.469$ & $2.813$ \\
\hline
$0.75$ & $0$ & $1.014\times 10^{2}$ & $115630$ & $0.968$ & $1.791\times 10^{1}$ & $5.046$ \\
& $0.1$ & $1.009\times 10^{2}$ & $115517$ & $0.968$ & $1.788\times 10^{1}$ & $5.042$ \\
& $0.5$ & $9.108\times 10^{1}$ & $112577$ & $0.982$ & $1.703\times 10^{1}$ & $4.942$ \\
& $1$ & $5.568\times 10^{1}$ & $99548$ & $1.100$ & $1.364\times 10^{1}$ & $4.542$ \\
\hline
$0.95$ & $0$ & $1.012\times 10^{3}$ & $205566$ & $0.569$ & $1.791\times 10^{1}$ & $5.046$ \\
& $0.1$ & $1.002\times 10^{3}$ & $205025$ & $0.570$ & $1.788\times 10^{1}$ & $5.042$ \\
& $0.5$ & $7.264\times 10^{2}$ & $189184$ & $0.607$ & $1.703\times 10^{1}$ & $4.942$ \\
& $1$ & $1.657\times 10^{2}$ & $99548$ & $0.946$ & $1.364\times 10^{1}$ & $4.542$ \\
\hline
$0.998$ & $0$ & $1.265\times 10^{4}$ & $386466$ & $0.331$ & $1.108\times 10^{2}$ & $8.527$ \\
& $0.1$ & $1.192\times 10^{4}$ & $380761$ & $0.334$ & $1.093\times 10^{2}$ & $8.513$ \\
& $0.5$ & $2.692\times 10^{3}$ & $262502$ & $0.494$ & $7.059\times 10^{1}$ & $8.165$ \\
& $1$ & $2.518\times 10^{2}$ & $145149$ & $0.919$ & $2.813\times 10^{1}$ & $6.187$ \\
\hline
\end{tabular}
\end{table*}

We shall consider a stellar sized wormhole of mass $15M_{\odot}$ as a central compact object with an accretion rate $\dot{M_{0}}\sim 10^{19}$ gm.sec$^{-1}$ and study the effect of different values of $\lambda$ in the spacetime described by RDSWH solution (20)-(21). We should ensure that $r_{\textmd{\scriptsize{ms}}}>r_{+}$, which is a necessary condition so that the accreting particles do not reach the throat and disappear into the other universe even for extreme limit $a_{\star}=1$. A special feature of the RDSWH is that it does satisfy this condition for $\lambda <10^{-3}$, the strong field lensing constraint for mimicking BHs \cite{Nandi:2018} (see Fig.1).

For illustration, we will consider different values of spin parameter $a_{\star}$. Choice $a_{\star}=0$ corresponds to static case of DSWH, which in turn goes to Schwarzschild solution when $\lambda = 0$. Choice $a_{\star}=0.75$ is expected from the collapse of a maximally rotating polytropic star \cite{Shapiro:2002}. Another highly probable value of spin parameter is $a_{\star}=0.95$, which comes from different observational methods \cite{Gou:2011,Walton:2016}. Case $a_{\star} = 0.998$ is a maximally allowed value coming from the Page-Thorne limit for Kerr BH \cite{Page:1974}.

\section{Numerical estimates}
\label{sec:5}
We present below three tables showing observable characteristics of the thin disk (Table 1), the minimum stable radius and the accretion efficiency (Table2) and the difference of the maxima of flux of radiation between RDSWH and Kerr BH (Table 3).

Figs.2a-2d display the flux of radiation $F(r)$ emitted by the disk between $r_{\textmd{\scriptsize{ms}}}$ and received at an arbitrary radius $r$ away from the center of the disk [Eq.(7)]. Figs.3a-3d show variation of temperature over the disk from $r_{\textmd{\scriptsize{ms}}}$ to an arbitrary radius and Figs.4a-4d show observed luminosity variations over different frequency ranges. The peaks values of emissivity properties of accretion disk are given in the Tab.1. In Tab.2, we show the variation of marginally stable orbits $r_{\textmd{\scriptsize{ms}}}$ and $\epsilon$ with the spin parameter $a_{\star}$ in the range used in the previous plots. The conversion efficiency of RDSWH is independent of $\lambda$, so is identical with that of Kerr BH.

From Figs 2a-2d we see that the flux emitted from the disk around Kerr BH always bigger than RDSWH depending on $a_{\star}$. To show the difference of profiles we will use their maxima given in Table 1. With $\lambda\leq 0.1$ flux of the accretion disk is practically indistinguishable from Kerr BH. Only limiting case $a_{\star}=0.998$ shows the difference of fluxes up to $6\;\%$. With the increase of the spin parameter the difference of fluxes of Kerr BH and RDSWH is also increasing. The most significant difference appears when $\lambda =1$, here we see that maximum value of flux of the wormhole is only $2\;\%$ of maximum value of Kerr BH flux. As $r\rightarrow\infty$ the fluxes of RDSWH and Kerr BH become indistinguishable. Another interesting feature of RDSWH is the shift of maxima of fluxes with the increase of $\lambda$ and spin parameter $a_{\star}$.

Similar characteristics appear in the disk temperature profiles, depicted in Figs. 3a-3d. For all values of the spin parameter the disks rotating around the Kerr black holes are hotter than those around the wormhole. With the increase of parameter $\lambda$ the temperature of the disk around rotating WH decreases.

In Figs. 4a-4d, we display the disk spectra for the spinning wormholes compared to Kerr BH. The cut-off frequencies for the Kerr black hole are systematically higher than those for the wormholes. The biggest deviations in radiation emitted by the accretion disk appears in ultraviolet range. Same as flux and temperature profiles the emissivity profile displays the most significant difference between RDSWH and Kerr BH when $a_{\star}=0.998$ and $\lambda = 1$.

\begin{table}[!ht]
\caption{The $r_{\textmd{\scriptsize{ms}}}$ and the efficiency $\epsilon$ for RDSWH. They are the same as those of Kerr BH. The general relativistic Schwarzschild black hole corresponds to $a_{\star}=0$. Here $a_{\star}=a/M$.}
\centering
\begin{tabular}{|c|c|c|}
\hline
$a_{\star}$ & $r_{\textmd{\scriptsize{ms}}}$ [$M$] & $\epsilon$ \\ \hline
$0$ & $6.00$ & $0.057$ \\
$0.3$ & $4.98$ & $0.069$ \\
$0.7$ & $3.39$ & $0.104$ \\
$0.92$ & $2.18$ & $0.205$ \\
$0.95$ & $1.94$ & $0.228$ \\
$0.998$ & $1.24$ & $0.321$ \\ \hline
\end{tabular}
\end{table}

Recently, by Bueno \emph{et al.} \cite{Bueno:2018} have considered small values $\lambda = 10^{-5}, 10^{-10}$ and $a_{\star}=0.7$, to analyze how the quasinormal modes of the RDSWH differ from those of the Kerr BH. If we consider the accretion disk properties, we can see, that only flux of radiation of wormhole is differ from BH. Others, temperature and emission spectra indistinguishable then $\lambda = 10^{-5}, 10^{-10}$.

In Table 3 below is shown the difference $\Delta_{\lambda}$ in the maxima of flux of radiation between RDSWH and Kerr BH:
\begin{equation}
\Delta_{\lambda} = F_{\textmd{\scriptsize{max}}}\big|_{\lambda = 0} - F_{\textmd{\scriptsize{max}}}\big|_{\lambda \neq 0}.
\end{equation}

\begin{table}[!ht]
\caption{Difference of the maxima of fluxes of radiation between RDSWH and Kerr BH, using $a_{\star}=0.7$.}
\centering
\begin{tabular}{|c|c|}
\hline
$\lambda$ & $\Delta_{\lambda}$[erg s$^{-1}$ cm$^{-2}$] \\ \hline
$10^{-5}$ & $2.54\times 10^{5}$ \\
$10^{-6}$ & $2.42\times 10^{3}$ \\
$10^{-7}$ & $28$ \\
$10^{-8}$ & $0$ \\
$10^{-9}$ & $0$ \\
$10^{-10}$ & $0$ \\ \hline
\end{tabular}
\end{table}

\section{Conclusions}
\label{concl}
Accretion around spinning wormholes is an active area of current research. Along this line, gravitational wave echoes of RDSWH have been recently studied by Bueno \textit{et al.}  \cite{Bueno:2018}. We studied in this work the thin accretion disk profiles of the same wormhole using the steady-state Page-Thorne model. Our aim was to examine how far the attractive positive mass mouth of RDSWH mimicks profiles of the Kerr BH for small values of $\lambda$ inspired by strong field lensing, which in general is an excellent diagnostic for comparison between two categories of objects (BH and WH) with differing topology, see. e.g., \cite{Nandi:2018,Izmailov:2019}. We chose for illustration a rotating stellar sized wormhole as the accreting object and obtained certain non-trivial generic characteristics of the disk. Specifically, we show that RDSWH parameter $\lambda$ has no influence on the kinematic profiles of the disc. That is, the accretion efficiency $\epsilon $, radius of the marginally stable orbit $r_{\textmd{\scriptsize{ms}}}$ (Table 2), the specific angular momentum $\widetilde{L}$, the specific energy $\widetilde{E}$, angular velocity $\Omega$ for a test particle in a circular equatorial orbit around RDSWH are \textit{independent} of $\lambda$.

However, differences appear in the emissivity properties for higher values $0.1<\lambda\leq 1$ (say) and for the extreme spin $a_{\star}=0.998$. We numerically computed the time averaged flux $F(r)$, temperature distribution $T(r)$ and emission spectra $\nu L(\nu )$ of the accretion disk for different $\lambda$ for the same dimensionless spin parameter $a_{\star}$. The relevant results are displayed in Figs. 2a-2d, 3a-3d, and 4a-4d for an accreting wormhole with assumed mass $15M_{\odot}$ and accretion rate $\dot{M_{0}}\sim 10^{19}$ gm.sec$^{-1}$ for $\lambda =0$ (Kerr BH), $\lambda =0.1$, $\lambda =0.5$ and $\lambda =1$ (RDSWH) respectively. Specifically, we observe certain characteristics of the emissivity profile, viz., the maxima (peak) of the $F(r)$, temperature distribution $T(r)$ and emission spectra $\nu L(\nu)$ shift in the following manner as can be read off from Table 1: For a fixed $a_{\star}$, an increment of $\lambda$ leads to an increasing shift in $r=r_{\textmd{\scriptsize{crit}}}$ at which the peaks of $F_{\textmd{\scriptsize{max}}}(r)$, $T_{\textmd{\scriptsize{max}}}(r)$ decrease, while there is a decreasing shift in the critical frequency $\nu_{\textmd{\scriptsize{crit}}}$ at which the peak of $\nu L(\nu)_{\textmd{\scriptsize{max}}}$ decreases. On the other hand, for a fixed $\lambda$, an increment of $a_{\star}$ leads to a decreasing shift in $r=r_{\textmd{\scriptsize{crit}}}$ at which the peaks of $F_{\textmd{\scriptsize{max}}}(r)$, $T_{\textmd{\scriptsize{max}}}(r)$ increase, while there is an increasing shift in the critical frequency $\nu_{\textmd{\scriptsize{crit}}}$ at which the peak of $\nu L(\nu)_{\textmd{\scriptsize{max}}}$ increases. These are generic features as variations of the wormhole mass and the rate of accretion within the RDSWH model preserve these characteristics. Thus the behavior of the luminosity peak is generically quite opposite to each other in the indicated variations, which in principle could be useful from the viewpoint of observations.

Apart from luminosity, it follows that an estimate of the difference $\Delta_{\lambda}$ in the maxima of flux of radiation $F(r)$ between RDSWH and Kerr BH shows non-zero values (Table 3) but is unobservable at present for small values $\lambda <10^{-3}$ permitted by the strong lensing bound. Given the fact that the accretion profiles are not known with great precision, the overall conclusion is that RDSWH are experimentally indistinguishable from KBH by accretion characteristics as of now. Nevertheless, microlensing perturbations to the flux ratios of gravitationally lensed quasar images can constrain the temperature profiles of their accretion disks (connected by Stefan-Boltzmann law), which could be a more accurate test for the accretion process \cite{Blackburne:2011} and may effectively lead to a tighter constraint on $\lambda$. Work is underway.


\begin{acknowledgments}
The reported study was funded by RFBR according to the research project No. 18-32-00377.
\end{acknowledgments}


\end{document}